
\documentstyle[bezier,12pt]{article}
\input pream1
\include{user:[valle.texfile]matex}
\include{matex}

\def\gsim{\;
\raise0.3ex\hbox{$>$\kern-0.75em\raise-1.1ex\hbox{$\sim$}}\;
}
\def\lsim{\;
\raise0.3ex\hbox{$<$\kern-0.75em\raise-1.1ex\hbox{$\sim$}}\;
}
\def\MeV{\rm MeV}
\def\eV{\rm eV}
\thispagestyle{empty}
\begin{titlepage}
\today
\begin{center}
\hfill FTUV/94-05\\
\hfill IFIC/94-02\\
\vskip 0.3cm
\LARGE
{\bf Nucleosynthesis constraints on active-sterile neutrino
conversions in the early universe with random magnetic field}
\end{center}
\normalsize
\vskip1cm
\begin{center}
{\bf V.B.Semikoz}
\footnote{On leave from the {\em Institute of the
Terrestrial Magnetism, the Ionosphere and Radio Wave Propagation of the
Russian Academy of Sciences, IZMIRAN,Troitsk, Moscow region, 142092, Russia}.}
{\bf and J.W.F. Valle}
\footnote{E-mail VALLE at vm.ci.uv.es or 16444::VALLE}\\
\end{center}
\begin{center}
\baselineskip=13pt
{\it Instituto de F\'{\i}sica Corpuscular - C.S.I.C.\\
Departament de F\'{\i}sica Te\`orica, Universitat de Val\`encia\\}
\baselineskip=12pt
{\it 46100 Burjassot, Val\`encia, SPAIN         }\\
\vglue 0.8cm
\end{center}

\begin{abstract}

We consider active-sterile neutrino conversions in the
early universe hot plasma in the presence of a random
magnetic field generated at the electroweak phase
transition. Within a random field domain the magnetization
asymmetry of the lepton antilepton plasma produced by
a uniform constant magnetic field is huge in contrast
to their small density asymmetry, leading to a drastic
change in the active-sterile conversion rates.
Assuming that the random field provides the
seed for the galactic field one can estimate
the restrictions from primordial nucleosynthesis.
Requiring that the extra sterile \neu does not
enter in equilibrium with the active ones before
nucleosynthesis we find limits of the oscillation
parameters which are stronger than in the isotropic
case.

\end{abstract}

\end{titlepage}

\newpage
\section{Introduction}
Recent observations of cosmic background temperature anisotropies
on large scales by the COBE satellite indicate the need for the
existence of a hot dark matter (HDM) component, contributing
about 30\% of the total mass density, i.e. $\Omega_{HDM}\sim
0.3$ \cite{Smoot}. Simple extensions of the standard electroweak
model that can reconcile all known hints for neutrino masses,
including solar and atmospheric neutrino observations
postulate the existence of a light sterile neutrino
$\nu_s$ \cite{Valle1,Valle2}.  In some of these models
such light sterile \neu is the HDM candidate \cite{Valle1}.

The most stringent constraints for the neutrino mass matrix
including such a fourth kind of neutrino, a singlet $\nu_s$,
are obtained from the nucleosynthesis bound on the
maximum number of neutrino species ($\Delta N \leq 0.3$)
that can reach thermal equilibrium before nucleosynthesis
and change the primordially produced helium abundance \cite{Schramm}.

In an isotropic early Universe hot plasma such constraint on the
additional neutrino species leads to an {\sl excluded} region in the
oscillation parameters $\mid\Delta m^2\mid$, $\sin^22\theta_0$
characterizing the active-sterile neutrino oscillations which
can be estimated (in non-resonant case) as \cite{Enqvist}
\footnote{
The discrepancy between these estimates and those refs.\cite{Kain,Barbieri}
stems  mainly from different estimates for the collision rate which
the authors of \cite{Enqvist} have evaluated in detail.
}
\barray{ll}[v1.1]
&\sin^42\theta_0\mid \Delta m^2\mid \gsim 5\times 10^{-6}eV^2,~~~\nu_a =
\nu_e\\
&\sin^42\theta_0\mid \Delta m^2\mid \gsim 3\times 10^{-6}eV^2,~~~\nu_a =
\nu_{\mu ,\tau }.\\
\ee

In this note we would like to reconsider the active-sterile neutrino
oscillation parameters assuming a new physical state of the
hot ultrarelativistic plasma before nucleosynthesis ($T \gg m_e$)
with the inclusion of the random magnetic field hypothesis
proposed in ref. \cite{Vachaspati}. This random magnetic
field could be generated at the electroweak phase transition
near the temperature $T \sim T_{EW}$ and could provide the
seed for the galactic field in the dynamo enhancement
mechanism \cite{Olesen}. In ref. \cite{Sem93,Shukurov} this
hypothesis was used in order to place stringent
constraints on the Dirac neutrino magnetic moments.

In this paper we neglect neutrino magnetic moments, both
diagonal as well as transition moments, and consider the
magnetization asymmetry of the primordial early universe hot plasma
produced by huge random magnetic fields. This influences
the neutrino spectrum in the medium and modifies the
neutrino conversions $\nu_a \leftrightarrow \nu_s$.

We confine ourselves to a small random magnetic field
domain size $L_0$, obeying the inequality
$L_0 \ll l_H$. Within such a domain the
magnetic field may be taken as uniform and constant,
so that the magnetization of the plasma can be easily
calculated. Here $l_H \sim M_{pl}/T^2$ is the horizon
length, $M_{pl}$ is the Planck mass and T is the plasma
temperature.

Although the magnetic field in different domains is
randomly aligned relative to the neutrino propagation
direction, we show how the observable neutrino conversion
probabilities depend on the mean-squared random field
via a squared magnetization value, therefore leading to
nonvanishing averages over the magnetic field distribution.
We apply this to the active-sterile neutrino conversions in
order to obtain more stringent limits than those that apply in
the absence of magnetic field.

\section{Dirac Neutrino Spectrum in a Hot Plasma with Magnetic Field}

In order to derive the Dirac equation for neutrino propagation
in a uniform locally anisotropic medium characterized by a
constant (within a domain) magnetic field ${\bf B}= (0,0,B)$
we start from the effective four-fermion interaction Lagrangian
of the standard electroweak theory
\be
{\cal L}_{int} = G_F \sqrt {2}\bar {\psi}_{\nu_b}\gamma^{\mu}{(1 -
\gamma_5)\over 2}\psi_{\nu_b}\sum_a\langle \bar {\psi}_a
[c^{ba}_V\gamma_{\mu} - c^{ba}_A
\gamma_{\mu}\gamma_5] \psi_a \rangle_0...,
\ee
where the indices $a$, $b = e, \mu, \tau$ correspond to the lepton
generations, and the dots denote the extra terms including hadrons.
Here $c^{ba}_V = 2 \sin^2\theta_W \pm 0.5$ is the vector
coupling constant (upper sign for $a = b$),
$\sin^2\theta_W$ is the electroweak mixing
parameter and $~c_A^{ba} = \mp 0.5$ is the
corresponding axial coupling constant
(upper sign for $b = a$). Finally, the symbol
$ \langle ...\rangle_0 $ denotes the statistical
averaging of vector and axial vector currents,
using equilibrium Fermi distributions
$f_{\kappa^{'}\kappa}^{(a)} =
 \langle b_{\kappa}^{(a)+}b_{\kappa'}^{(a)}\rangle_0 =
f^{(a)}_{\lambda^{'}\lambda }(p_z, n)$. Explicitly,
these are given as
\be[v1.9]
f_{\kappa^{'}\kappa}^{(a)} = \frac{\delta_{\lambda \lambda^{'}}}
{\exp [(\varepsilon_{n\lambda}(p_z) - \zeta_a)/T] + 1},
\ee
 where the lepton spectrum is of the form
\be[v1.10]
\varepsilon_{n\lambda}(p_z) = \sqrt{p_z^2 + m^2_a + \mid e\mid B(2n + 1 -
\lambda)}.
\ee

For the case of antileptons, one has

$f_{\kappa^{'}\kappa}^{(\tilde {a})} =
 \langle d_{\kappa}^{(a)+}d_{\kappa'}^{(a)}\rangle_0 =
f^{(a)}_{-\lambda^{'},-\lambda }(p_z,n)$, or explicitly,
\be[v1.9a]
f_{\kappa^{'}\kappa}^{(\tilde {a})} = \frac{\delta_{-\lambda,-\lambda^{'}}}
{\exp [(\varepsilon_{n\lambda}(p_z) + \zeta_a )/T] + 1},
\ee
 where the antilepton spectrum is of the form
\be[v1.10a]
\varepsilon_{n\lambda}(p_z) = \sqrt{p_z^2 + m^2_a + \mid e\mid B(2n + 1 +
\lambda)}.
\ee
Here $\zeta_a$ is the chemical potential and the full set
of the quantum numbers $\kappa$ includes $\{p_z,n,\lambda\}$,
where $p_z$ is the conserved momentum component for the chosen
magnetic field geometry; $n = 0,1,2,...$ is the Landau number
and $\lambda =\pm 1$ is twice the eigenvalue of
the conserved lepton spin projection on the magnetic
field, $(\sigma_z)_{\lambda'\lambda} = \lambda \delta_{\lambda'\lambda}$.
The change of sign $\lambda$ in \eq{v1.9a} arises from the conjugation
property $C\sigma_iC^{-1} = - \sigma_i^T$.

The resulting equation describing the neutrino motion
takes the form
\barray{ll}[v1.2]
\Bigl [\hat{q} - m^{vac}_{\nu_b} - V^{(vec)}_b\gamma_0\frac{(1 - \gamma_5)}
{2} - V_b^{(axial)}\gamma_z\frac{(1 - \gamma_5)}{2}\Bigr ]\Bigl (\matrix{
\varphi \\ \chi }\Bigr ) = 0,
\ee
where the vector interaction potential of an active (left) neutrino
$\nu_b$ ($b=e,\mu, \tau$), $V_b^{(vec)} = G_F \sqrt {2}\sum_ac^{(ab)}_V
\langle \bar {\psi}_a\gamma_0\psi_a\rangle$
\footnote{
Note that spatial components of the mean vector current are zero,
$\langle \bar {\psi}\gamma_i\psi\rangle = 0$, $i = 1,2,3$.}
is given by the known formula \cite{Enqvist}:
\be[v1.3]
V_b^{(vec)} = G_F \sqrt{2} n_{\gamma}[L_0^{(\nu_b)} - A_b\frac{T^2}{M_W^2}].
\ee
The first term in the vector potential \eq{v1.3} that is proportional to
the small particle-antiparticle asymmetries
$L_{\alpha} = (n_{\alpha} - n_{\tilde {\alpha}})/n_{\gamma}$,
normalized on the photon density $n_{\gamma}=0.244T^3$,
is given by  \cite{Enqvist}
\be
L_0^{(\nu_b)} = [\sum_ac_V^{ba}\times L_a +
(1 - 4 \sin^2\theta_W )L_p/2 - L_n/2 + 2L_{\nu_b} + \sum_{a\neq b}L_{\nu_a}].
\ee
This term is changed a little due to the effect of the strong
magnetic field on the charged lepton (antilepton) densities.
However this change remains negligible compared with the
non-local second term in \eq{v1.3} which, for the case of
electron neutrinos, is given by \cite{Notzold}
\be[v1.4]
\mid V^{(vec)}_{\nu_e}\mid \approx 3.4\times 10^{-20}\Bigl (\frac{T}{\MeV}
\Bigr )^5~\MeV.
\ee
We have neglected the influence of the
magnetic field on the nucleon densities, due to the very
small magnetic moments of the nucleons and the small
temperatures below the QCD phase transition ($T\lsim 200~\MeV$).

Now we turn to the new axial term in the Dirac equation, namely
\be[v1.5]
V_b^{(axial)} = \frac{G_F}{\sqrt{2}}\sum_{a=e,\mu ,\tau}(-2c^{ba}_A)
\langle \bar{\psi }_a\gamma_z
\gamma_5\psi_a\rangle_0
\ee
This term corresponds to the magnetization
asymmetry of each component ($a=e, \mu, \tau$)
of the hot plasma in the external magnetic field, $M_z^{(a)} - M_z^{(\tilde
{a})}$,
\be[v1.6]
M_i^{(a)} - M_i^{(\tilde {a})} =
\mu_{B}\langle \psi_a^+
\Sigma_i\psi_a\rangle_0 \equiv \mu_B\langle \bar{\psi }_a\gamma_i
\gamma_5\psi_a\rangle_0,
\ee
where $\mu_B = \mid e\mid /2m_a$ is the Bohr magneton.
Note that for our chosen magnetic field geometry
only the z-component of the axial current is nonvanishing.

The new macroscopic axial term $V^{(axial)}_b$
changes the active neutrino spectrum in the plasma.
In the ultrarelativistic limit $m_{\nu_b} \to 0$
such spectrum can be obtained, from \eq{v1.2}, as
\be[v1.7]
E = V_b^{(vec)} + \sqrt{q^2_{\perp} + (q_z + V_b^{(axial)})^2},
\ee
This differs from the isotropic one,
$E = V_b^{(vec)} + q$, due to the shift of the
z-component of the total neutrino momentum $q = \sqrt{q_{\perp}^2 +
q_z^2}$. In the hot plasma the neutrino momentum $\langle q\rangle
\sim 3 T$ is larger than both $V_b^{(vec)}$ and $V_b^{(axial)}$,
so that we can use for neutrino oscillation problem
the ultrarelativistic approximation,
\be[v1.8]
E \approx q + V_b^{(vec)} + V_b^{(axial)}\frac{q_z}{q} + \frac
{(V_b^{(axial)})^2q^2_{\perp}}{2q^3}.
\ee

\section{Magnetization Asymmetry of Hot Plasma
	in an External Magnetic Field}

In order to show the importance of the axial term contribution \eq{v1.5}
in the neutrino spectrum, let us consider the main features of
the hot lepton-antilepton plasma in a uniform constant magnetic
field ${\bf B} = (0,0,B)$.

One can easily check that the lepton-antilepton density asymmetry
$\langle \bar {\psi}_a\gamma_0 \psi_a\rangle_0 = n_a - n_{\tilde{a}}$
equals to
\be[v1.11]
n_a - n_{\tilde{a}} = \sum^{\infty }_{n = 0}\frac{\mid e \mid B}{(2\pi )^2}
\int^{\infty }_{-\infty }dp_zTr[f^{(a)}(p_z,n) - f^{(\tilde{a})}(p_z,n)],
\ee
where trace is calculated over spin variables $\lambda $
and the (--) sign inside \eq{v1.11} arises from the N-ordering
of operators in the current $N(\bar{\psi}_a \gamma_{\mu } \psi_a)$.

In order to check the normalization of the Fermi-distribution \eq{v1.9},
let us consider the WKB approximation $n \gg 1$ in the weak magnetic
field limit $\mid e\mid B \ll T^2$ changing $2\mid e \mid Bn$
into $p_{\perp }^2$. Using $\mid e\mid B dn = p_{\perp }dp_{\perp }$
one obtains the standard isotropic result
\be
n_a - n_{\tilde{a}} = 2\int\frac{d^3p}{(2\pi)^3}[\frac{1}{\exp
((\sqrt{p^2_z
+ p^2_{\perp } + m^2_a} - \zeta )/T + 1)} - (\zeta \to -\zeta )],
\ee
for the lepton-antilepton asymmetry \eq{v1.11}. The factor
"2" is produced by the spin sum. Note that the contribution
of our general asymmetry \eq{v1.11} to the neutrino vector
potential
\eq{v1.3} is negligible comparing with the main non-local term \eq{v1.4}.

Now we calculate with help of \eq{v1.9} the magnetization
asymmetry in the hot plasma. In analogy with \eq{v1.11}
we can write the lepton contribution to the magnetization
asymmetry as
\be[v1.12]
M_j^{(a)} = \mu_B\sum^{\infty }_{n = 0}\frac{\mid e\mid B}{(2\pi)^2}
\int^{\infty }_{-\infty }dp_zTr[\sigma_jf^{(a)}(p_z,n)].
\ee
Using the trace $Tr[\sigma_jf^{(a)}] = \sum_{\lambda \lambda'}
(\sigma_j)_{\lambda \lambda^{'}}f^{(a)}_{\lambda^{'}\lambda }(p_z,n) =
\delta_{jz}\sum_{\lambda }\lambda f^{(a)}_{\lambda \lambda }(p_z,n)$
with $(\sigma_z)_{\lambda \lambda^{'}} = \lambda \delta_{\lambda
\lambda^{'}}$ and $\delta_{\lambda \lambda } = 1$ for $\lambda = \pm 1$,
one can easily show that, due to the degeneracy of
the Landau levels $n = 1,2,...$,
$\varepsilon_{n+1,1} = \varepsilon_{n,-1}$)
all terms in the sum
\be
\sum_{n=0}^{\infty} \frac{1}{\exp ((\varepsilon_{n,1} - \zeta)/T) + 1} -
\frac{1}{\exp((
\varepsilon_{n,-1} - \zeta )/T) + 1},
\ee
cancel
\footnote{One can easily check that in the weak magnetic field
limit $\mid e\mid B\ll T^2$, due to the subtraction of the
Fermi distribution functions with different spin projections,
from the relativistic magnetization \eq{v1.12} one recovers
the known result corresponding to the spin paramagnetism of
the non-relativistic free electron gas in a metal \cite{Kubo}:
\be
M_z = - 2\mu_B^2B\int D(E){d~f\over dE}dE .
\ee
Note that here we used the isotropic phase volume with the energy
$E = p^2/2m_e$, $D(E) = (2m_e)^{3/2}\sqrt{E}/(2 \pi)^2$ and that
the Fermi-distribution in non-relativistic case is given by
$f(E) = [\exp ((E - \zeta')/T) + 1]^{-1}$ where the chemical
potential is $\zeta' = \zeta - m_e$.}
except for contribution of the
main (non-degenerate) Landau level ($n = 0$)
\be
\frac{1}{\exp ((\varepsilon_{0,1} - \zeta )/T) + 1} =
\frac{1}{\exp ((\sqrt{p_z^2 +
m_a^2} - \zeta )/T) + 1}.
\ee

Finally, the lepton part becomes
\be[v1.13]
M_z^{(a)} = \mu_B \frac{2\mid e\mid B}{(2\pi )^2}\int_0^{\infty }dp_z[\frac{1}{
\exp((\sqrt{p^2_z + m^2_a} - \zeta )/T) + 1}].
\ee

Now we turn to the antilepton part. Using the C-conjugation
of the relevant operators we must add to the trace in the
integrand of \eq{v1.12} the trace over negative spin projections
\be[v]
- \sum_{\lambda ,\lambda'}(-\sigma_j^T)_{-\lambda';-\lambda}
f^{(\tilde{a})}_{-\lambda;-\lambda'}(p_z,n),
\ee
where the (--) sign before the sum arises from the N-ordering
of the operators in the axial current
$N(\bar {\psi}_a\gamma_{\mu}\gamma_5\psi_a)$, and
the second one comes from the C-conjugation property
$C\sigma_jC^{-1}= - \sigma_j^T$. Now one can easily
show, by using the change $\lambda \to -\lambda$,
that the antilepton part may be rewritten as
\be[v1.15]
- \sum_{\lambda \lambda'}(-\sigma_j)_{\lambda';\lambda}f^{(\tilde{a})}_
{\lambda\lambda'}(p_z,n)
= - \sum_{\lambda\lambda'}\frac{(-\lambda)\delta_{\lambda'\lambda}
\delta_{\lambda
\lambda'}}{\exp ((\sqrt{p_z^2 + m_a^2 + \mid e\mid B(2n + 1 - \lambda)}
 + \zeta)/T) + 1},
\ee
where we used the conserved spin z-component eigenvalue
$(\sigma_z)_{\lambda'\lambda} = + \lambda\delta_{\lambda'\lambda}$.
Similarly as obtained above for the lepton part, all of the
Landau-level contributions $n = 1,2...$ cancel due to
the degeneracy property
$\varepsilon_{n + 1,1} = \varepsilon_{n, -1}$.
Thus the magnetization asymmetry is obtained
by {\sl summing} lepton and antilepton contributions,
in contrast to the density asymmetry,
\bea
M_z^{(a)} - M_z^{(\tilde{a})}=
\mu_B \frac{2\mid e\mid B}{(2\pi )^2}
\int_0^{\infty }dp_z[{1\over
\exp((\sqrt{p^2_z + m^2_a} - \zeta )/T) + 1}  \\\nonumber
+ \frac{1}{\exp((\sqrt{
p_z^2 + m^2_a} + \zeta )/T) + 1} ].
\eea
Substituting this result into the axial potential in \eq{v1.5}
and \eq{v1.6} one can finally rewrite the ultrarelativistic
($m_a = \zeta = 0$) neutrino spectrum \eq{v1.8} as
\be[spectrum]
E = q + V_b^{(vec)} + f(q,B),
\ee
where the magnetic field contribution $f(q,B) = V^{(axial)}_bq_z/q +
(V_b^{(axial)})^2q^2_{\perp}/2q^3$ is given by
\be[v1.18]
f(q,B) = \mu_{eff} \frac{{\bf qB}}{q} + \frac{\mu^2_{eff}}{2q}\Bigl (B^2 -
\frac{({\bf qB})^2}{q^2}\Bigr ).
\ee
and the quantity  $\mu_{eff}$ is defined by
\footnote{Note that this effective magnetic moment
has no relation with the real anomalous neutrino
magnetic moment which we neglect.}
\be[v1.19]
\mu_{eff} = \frac{e G_F (-2c_A) T \ln 2}{\sqrt{2}\pi^2} \approx 6 \times
10^{-13} \mu_B (\frac{T}{\MeV}).
\ee
For a hot plasma ($T \gg 1~\MeV$) this is huge, in contrast
to the small lepton-antilepton density asymmetry \eq{v1.11}.
This arises from the fact that the magnetization
asymmetry is produced by the mean $axial$ current
so that the lepton and anti lepton contributions
$add$ instead of subtract. In a strong uniform
magnetic field the first term in \eq{v1.18} may
exceed the non-local term \eq{v1.4} considered
by Notzold and Raffelt.
As we will show below, for large random magnetic
fields, such term can drastically change the
active-sterile neutrino conversion rates.

\section{Active-sterile neutrino conversions in
a hot plasma with random magnetic field}

Let us now consider the wave equation describing the
propagation of a system of active (doublet) and
light sterile (singlet) neutrinos, with masses
$m_1$ and $m_2$, mixing angle $\theta$, and no
transition magnetic moments, in the presence of
a random magnetic field.
We postulate the following evolution equation
\footnote{Strictly speaking, in order to describe
the active to sterile \neu conversions one has to
start from a system of two majorana \neus and not
from the Dirac equation as we did in section 2
\eq{v1.2}. This can be done and one finds
\cite{sergio} that our ansatz is obtained
in the ultrarelativistic small mixing angle limit. }:
\barray{ll}[v1.20]
i\frac{d}{dt}\Bigl (\matrix{ \nu_d \\ \nu_s }\Bigr ) = \Bigl [
\matrix{ (c^2m^2_1 + s^2m^2_2)/2q + V_d + f(q, B) & c s \Delta \\
 s c \Delta & (s^2m^2_1 + c^2m^2_2)/2q}\Bigr ] \Bigl (\matrix{ \nu_d \\
\nu_s }\Bigr ),
\ee
where we use the standard definitions
$\Delta = \Delta m^2/2q$;
$\Delta m^2 = m^2_2 - m^2_1$;
$c = \cos \theta$, and
$s = \sin \theta $. In addition we have denoted
by $V_d$ the vector part of the active neutrino
potential of \eq{v1.3}.

In the phase of the early universe hot plasma of
interest to us we have only active neutrinos in thermal
equilibrium. From \eq{v1.20} one can easily obtain a
nonlinear integro-differential equation for the
conversion probability $P_{\nu_d \to \nu_s} (t)$
from active to sterile neutrinos, $\nu_d \rightarrow \nu_s$ .
It obeys the unitarity condition
\be
P_{\nu_d \to \nu_s} (t) = \VEV{\nu^*_s\nu_s} =
1 - P_{\nu_d \to \nu_d} (t) =
\langle \nu^*_d\nu_d\rangle ,
\ee
Defining $P_{\nu_d \to \nu_s} \equiv P (t)$ and
averaging over the ensemble of random magnetic
fields one obtains:
\be[v1.21]
\frac{d^2P}{dt^2} + \Delta^2_mP + \langle \Delta^2_B\rangle \int^t_0dt_1
K(t - t_1)\frac{dP(t_1)}{dt_1} = \frac{\Delta^2\sin^22\theta}{2},
\ee
where the initial conditions are given by
\be[v1.22]
P(0) = \dot{P}(0) = 0.
\ee
The factor before the second term,
\be[v1.23]
\Delta^2_m = (V_d - \Delta\cos 2\theta)^2 + \Delta^2\sin^22\theta
\ee
is the well known oscillation squared frequency in an isotropic
hot plasma \cite{Enqvist}. Due to the property of randomness
$\langle B_z\rangle = 0$ the factor before integral term in \eq{v1.21},
$\langle \Delta^2_B\rangle = 2(V_d - \Delta\cos2\theta)\langle f(q,B)
\rangle + \langle f^2(q,B)\rangle$, with the
function $f(q,B)$ from \eq{v1.18}, is determined mainly by the second
term $\langle \Delta^2_B\rangle \approx \langle f^2(q,B)\rangle $.
For collisionless neutrino propagation along z-axis
${\bf q} = (0,0,q)$, the factor $\langle \Delta^2_B\rangle $
takes of the form
\be[v1.24]
\langle \Delta^2_B\rangle \simeq \mu_{eff}^2\langle {\bf B}^2\rangle/3.
\ee
after averaging over random magnetic fields.

Here the effective neutrino magnetic moment is given by \eq{v1.19}
with the mean squared first term of \eq{v1.18} given by
$\langle B^2_z(t)\rangle =\langle {\bf B}^2\rangle/3$.
Note that the second term contribution can be neglected
in the ultrarelativistic limit. Finally the kernel
in \eq{v1.21}
$K(t_1 - t_2) =\langle B_z(t_1)B_z(t_2)\rangle/\langle B_z^2(t)\rangle$
depends on the model of random fields. If we choose the simple model
with uncorrelated magnetic field domains of the same small size $L_0$
\footnote{For neutrinos crossing many domains $t=L \gg L_0$
the size $L_0$ corresponds to the width of the narrow resonance
$L_0^2/(L_0^2 + t^2)$ for our assumed $\delta$-correlated random
fields
\be
\frac{K(t)}{L_0} \sim \lim_{L_0 \to 0}\frac{L_0}{t^2 + L_0^2} = \frac{\pi}{2}
\delta (t).
\ee
}
, $K(t_1 - t_2) = L_0 \delta (t_1 - t_2)$, the
integro-differential equation \eq{v1.21} reduces to a second
order differential equation with the boundary conditions
\eq{v1.22}.
The solution of this equation is of the form
\be[v1.27]
P(t) = \frac{\Delta^2\sin^22\theta}{2\Delta^2_m}\{1 - \exp (-\Gamma t)
\Bigl [\cosh (\sqrt{\Gamma^2 - \Delta^2_m}t) + \frac{\Gamma}{\sqrt{
\Gamma^2 - \Delta^2_m}}\sinh (\sqrt{\Gamma^2 - \Delta^2_m}t)\Bigr ]\},
\ee
where
\be[v1.28]
\Gamma = \frac{\langle \Delta^2_B\rangle L_0}{2}
\ee
is the damping parameter of neutrino oscillations in a random field.
In the weak magnetic field limit $\Gamma \ll \Delta_m$, where the neutrino
oscillation frequency in matter $\Delta_m$ is given by \eq{v1.23}, such
solution,
\be[v1.29]
P(t) \simeq \frac{\Delta^2\sin^22\theta}{2\Delta^2_m}(1 - \exp
(- \Gamma t)\cos \Delta_mt),
\ee
reproduces the known case of $\nu_d\leftrightarrow \nu_s$
oscillations in an isotropic hot plasma if $\Gamma = 0$.
For strong random magnetic fields generated at the
electroweak phase transition \cite{Vachaspati}
the opposite condition $\Gamma\gg \Delta_m$ is fulfilled.
The corresponding asymptotics of the solution \eq{v1.27},
\be[v1.30]
P(t) \simeq \frac{\Delta^2\sin^22\theta}{2\Delta^2_m}\Bigl (1 -
\exp (-\Delta^2_mt/2\Gamma )\Bigr ),
\ee
is mostly aperiodic, in contrast to \eq{v1.29}. From
\eq{v1.30} one define the relaxation time
$t_{relax}\sim 2\Gamma/\Delta_m^2 = \langle
\Delta^2_B\rangle L_0/\Delta_m^2$.
The condition that the \neu crosses many domains,
leads to the requirement
$\langle \Delta^2_B \rangle \gg \Delta_m^2$.
In the next section we verify that this
condition is indeed fulfilled.

\section{Random magnetic fields before nucleosynthesis}

In order to show the validity of \eq{v1.30} let us estimate
the factor $\langle \Delta^2_B\rangle$ in \eq{v1.24} in formula \eq{v1.28}
substituting to \eq{v1.24} the mean squared random magnetic field
\be[v1.31]
\sqrt {\langle B^2\rangle} = 10^{24}~G\Bigl (\frac{T}{T_{EW}}\Bigr )^2\times
\Bigl (\frac{L_0}{L}\Bigr )^p
\ee
with the scale dependence obeying the index $p=1/2$ \cite{Olesen93}.

Requiring that the primordial magnetic field survives
beyond the recombination epoch leads to a minimal domain
size \cite{Schramm1},
\be[v1.32]
L_0\gsim 10^4~cm(T_{BBN}/T) \sim 10^3 \rm{cm} (\MeV/T).
\ee
With this assumption let us now estimate a lower
limit for \eq{v1.24}. In order to do this we use
the collisionless neutrino propagation approximation,
i.e. $t=L\leq t_{coll}=\Gamma_W^{-1}$, and substitute
\eq{v1.32} into \eq{v1.31} leading to
\be[v1.33]
\langle \Delta^2_B\rangle^{1/2} =
\frac{\mu_{eff}\langle B^2\rangle^{1/2}}
{\sqrt{3}}\gsim 6\times 10^{-17}\Bigl (\frac{T}{MeV}\Bigr )^5 \MeV ,
\ee
Note that this is significantly larger than the nonlocal vector
potential term given in \eq{v1.4}. Therefore, we have obtained a
self-consistent requirement
$\langle \Delta_B^2\rangle \gg \Delta_m^2$ which justifies our
use of the $\delta$-correlated random magnetic fields.

If instead we put into \eq{v1.31} the maximum scale $L=l_H(T)$,
where $l_H$ is the horizon length at temperature T we obtain
another estimate
$\langle \Delta^2_B\rangle^{1/2}\gsim 2\times 10^{-16}(T/MeV)^{3.5}~MeV$,
which also exceeds \eq{v1.4}. However, in such case one should
include the effect of collisions on the \neu conversions.

Substituting the limit \eq{v1.32} rewritten as
$L_0\gsim (10^{14}/1.9)(MeV/T)~MeV^{-1}$ and the estimate
in \eq{v1.33} into \eq{v1.28} we see that random magnetic
field \eq{v1.31} obeys the condition
$\Gamma \gg \Delta_m\sim \mid V_d\mid$. This demonstrates
the validity of our main approximation to the conversion
probability \eq{v1.30}.

\section{Nucleosynthesis bounds on the sterile neutrino conversions}

Comparing the relaxation time in the probability \eq{v1.30}
estimated with help of \eq{v1.4}, \eq{v1.23}, \eq{v1.32} and \eq{v1.33}
as
\beq
t_{relax} \sim 2 \Gamma/ \Delta_m^2 \gsim 2 \times 10^{20}T^{-1}
\eeq
with the collision time,
\beq
t_{coll}\sim 2\times 10^{21}(MeV)^4T^{-5},
\eeq
one finds a critical temperature which separates
two regimes $T_* \sim 10^{1/4}~MeV$ above which
it should be important to take into account
collisions.
For times less than the neutrino collision time
$t_{coll} = \Gamma_W^{-1}$ we can directly estimate
the sterile neutrino conversion probability of
\eq{v1.30} in the collisionless approximation.
It is also very simple to consider the alternative
limit where one can average over many collisions.
In both cases one obtains essentially the same result.
Finally, for the case of intermediate temperatures
close to $T_* \sim 2$ MeV one needs a more accurate
kinetic approach as in ref. \cite{Enqvist}.

For definiteness, we consider here the regime the
derivation of the nucleosynthesis bounds for active-sterile
\neu oscillations in the case where one can average over many
collisions. Averaging $t_{coll}^{-1} \int_0^{tcoll}dtP(t)$
from \eq{v1.30} one obtains the result
\be[r1]
\langle P \rangle_{coll} = \frac{\Delta^2 \sin^2 2\theta }{2\Delta^2_m}
\Bigl [1 - \frac{2\Gamma \Gamma_W}{\Delta^2_m} + \frac{2\Gamma \Gamma_W}
{\Delta^2_m}\exp \Bigl (-\frac{\Delta^2_m}{2\Gamma \Gamma_W}\Bigr )
\Bigr ].
\ee
The factor in brackets above, $f(x) = 1 - x^{-1} + \exp(-x)/x$,
where $x = \Delta^2_m/2\Gamma \Gamma_W$, is a monotonic function
which attains its minimum value $f(0) = 0$ as $\Gamma \to \infty $
and tends asymptotically to $f(x = \infty ) = 1$ when one neglects
the magnetic field, i.e. $\Gamma \to 0$.
In a strong random magnetic field this factor is restricted by
$f(x_{max})$, when we substitute the estimates \eq{v1.32} and \eq{v1.33}
into the damping factor $\Gamma $ in \eq{v1.28},
\beq
\Gamma \geq \Gamma_{min} \approx 10^{-19}(T/MeV)^9~MeV
\eeq
or
\be[r2]
x \lsim x_{max} = \frac{\Delta^2_m}{2\Gamma_W\Gamma_{min}} \approx
10\Bigl (\frac{T}{\MeV }\Bigr )^{-4}
\ee
and $f(x) \lsim f(x_{max})$. In \eq{r2} we have used the neutrino
non-resonant oscillation frequency estimate $\Delta_m \sim V_d$,
where $V_d$ is given by the N\"{o}tzold-Raffelt result \eq{v1.4}
and the usual weak interaction rate
$\Gamma_W = 4.0 G_F^2 T^5 \sim 5.4 \times 10^{-22} (T/\MeV )^5~ \MeV$.

One can easily see from \eq{r2} that for temperatures
$T \gsim 3 \MeV $ the argument $x$ is very small, $x \ll 1$,
so we can rewrite the probability \eq{r1} as
\be[r3]
\langle P \rangle_{coll} = \frac{\sin^2 2 \theta_B}{4 L_0 \Gamma_W},
\ee
where the mixing angle in the the hot plasma with magnetic field
is given by
$\sin^22\theta_B = \Delta^2\sin^22\theta/\langle \Delta^2_B\rangle$
and is restricted due to \eq{v1.33} ($\langle q\rangle\sim 3T$)
by
\be[r4]
\sin^2 2\theta_B \simeq
\frac{\sin^2 2 \theta (\Delta m^2)^2}{36T^2
\langle \Delta^2_B\rangle}\lsim 10^7\Bigl (\frac{\Delta m^2}{\eV^2}\Bigr )^2
\Bigl (\frac{T}{\MeV}\Bigr )^{-12}\sin^22\theta.
\ee
Taking into account eq.(5.2), we find that the sterile neutrino
production rate
$\Gamma_{\nu_s} (t\leq t_{coll}) = P \Gamma_W$
obtained from \eq{r3},
$\Gamma_{\nu_s} = \sin^22\theta_B/4L_0$,
obeys the inequality
\be[r5]
\Gamma_{\nu_s} \lsim \frac{1}{2}
\times 10^{-7}\Bigl (\frac{\Delta m^2}{\eV^2}\Bigr )^2\sin^22\theta
\Bigl (\frac{T}{\MeV}\Bigr )^{-11}~\MeV.
\ee
Sterile neutrinos would be thermalized if this rate \eq{r5}
exceeds the Hubble expansion rate
$H = 4.5 \times 10^{-22} (T/\MeV)^2 \MeV$,
\be
\Gamma_{\nu_s}/ H \geq 1.
\ee
Using the inequality \eq{r5} we obtain a new constraint
on the $\mid \Delta m^2\mid, \sin 2 \theta $ oscillation
parameters,
\be[r6]
\mid \Delta m^2 \mid \abs{\sin2\theta} \lsim 10^{-7} \times
\Bigl (\frac{T}{\MeV}\Bigr )^{13/2} \eV^2.
\ee
Note that the sign of $\Delta m^2$ for us here is
irrelevant, in contrast to the isotropic case.

Self-consistency of our approximations (see \eq{r2})
requires us to assume in \eq{r6} a minimal temperature
$T \gsim T_{min} \sim 3~\MeV$ which then allows us to
rewrite \eq{r6} as
\be[r7]
\mid \Delta m^2\mid \abs{\sin 2\theta} \lsim 10^{-4}\eV^2.
\ee
We see that this bound can be significantly stronger
than the nonresonant estimate \eq{v1.1} obtained for the
case of an isotropic hot plasma (see Fig.1). This is especially so
for the case of small mixing angle
($\sin2\theta \sim 0.1$), where we obtain from \eq{r7}
an excluded region of the squared mass difference
$\mid \Delta m^2\mid \gsim 10^{-3} \eV^2$ instead
of the result of \eq{v1.1},
$\mid \Delta m^2\mid \gsim 5 \times 10^{-2} \eV^2$.

Note also that, in the case of the MSW resonance,
($\Delta^2_m = \Delta^2 \sin^22\theta $, see \eq{v1.23},
the argument $x = \Delta^2_m/2\Gamma
\Gamma_W$ in the probability \eq{r1} is less than
\be[r8]
x \lsim x_{max} = 3\times 10^{14}\Bigl (\frac{\Delta m^2}{\eV^2}\Bigr )^2
\sin^22\theta \Bigl (\frac{T}{\MeV }\Bigr )^{-16}.
\ee
If the upper limit here is much less than unity, we automatically
obtain from \eq{r1} the same probability \eq{r3} that does not
depend on the frequency $\Delta_m$ at all. Therefore, the
constraint \eq{r6} would be a general one and substituting
acceptable values of the product $(\Delta m^2/\eV^2)^2\sin^22\theta
\lsim 10^{-14}(T/\MeV)^{13}$ into \eq{r8} we confirm the validity
of the ultrarelativistic approximation
$x \lsim x_{max} \lsim 3\times (T/\MeV)^{-3} \ll 1$
for this resonant regime of $\nu_e \leftrightarrow \nu_s$
oscillations too.

\section{Discussion and conclusions}

The existence of huge magnetic fields generated at the
electroweak phase transition modifies the
\neu spectrum in the early universe hot plasma.
This happens due to the magnetization of each
component of the plasma. Within a small random magnetic
field domain the local uniform constant magnetic field
approximation allows us to calculate the magnetization,
which is proportional to the random magnetic field.

In contrast to the small lepton antilepton density
asymmetry produced by the mean vector current the
magnetization asymmetry produced by the mean axial current
is large, due to their opposite charge conjugation properties.
As a result the lepton and anti lepton magnetizations
$add$ instead of subtract.

Averaging the differential equation describing
the evolution of the $ \nu_d \to \nu_s$ conversion
probabilities over the random magnetic field distribution
we find a nonvanishing mean squared field contribution
which drastically changes the conversion rates
with respect to those of the isotropic case.

Assuming that the primordial magnetic fields generated
at the electroweak phase transition are the origin
of the observed galactic fields and requiring that there should
be no more than one extra \neu species in equilibrium before
nucleosynthesis we derive new and more stringent constraints
on the active-sterile \neu oscillation parameters than in the
isotropic case without random magnetic field. In contrast to
what happens in the isotropic case, our constraints do not
depend on the active-sterile \neu conversion channel beyond
the obvious dependence contained in $\Delta m^2$.

The fact that the constraints can be stronger in
our case than in the isotropic case, despite the
fact that the conversion rates smaller, follows
from the different way in which they depend on
the oscillation parameters. In particular, in the
random field case there is a more sensitive dependence
of the average conversion rates upon the \neu squared mass
difference. While in the isotropic case there is a
saturation of this probability as a function of the
\neu squared mass difference, in our case we have a
linear dependence upon $\Delta m^2$.

\vfill
\noi
{\bf Acknowledgements}
This work was supported by DGICYT under grant number
PB92-0084 and by a senior researcher NATO fellowship
(V.S.). We thank Dr. A. I. Rez who pointed out the
simple solution of \eq{v1.21}.

\vskip 1truecm

\newpage
\begin{center}
\vskip5cm
{\bf Figure Caption}
\end{center}
\vskip2cm
{\bf Fig.1.} \\
Regions of $\nu_e \leftrightarrow \nu_s$ oscillation
parameters excluded by nucleosynthesis:
(a) the region above the dotted line is excluded by
the requirement that $N_{\nu } < 3.4$ for the non-resonant
isotropic hot plasma estimate of ref. \cite{Enqvist};
(b) the region above the solid line is excluded in present work
for the the case a hot plasma with primordial random magnetic
field, seed for the galactic field, by the requirement that
$N_{\nu } < 4$ for $T \gsim T_{min} \sim 3\MeV $, \eq{r7}.

\end{document}